\documentclass[12pt,a4paper]{article}
\usepackage{latexsym}

\def\nn{\nonumber}
\def\K1{{\cal K}_{\bf 1}}
\def\Q20{{\cal Q}_{\bf 20'}}

\newcommand{\ie}{{\em i.e.~}}
\newcommand{\eg}{{\em e.g.~}}

\newcommand{\be}{\begin{equation}}
\newcommand{\ee}{\end{equation}}
\newcommand{\ba}{\begin{eqnarray}}
\newcommand{\ea}{\end{eqnarray}}

\begin{document}

\begin{titlepage}
\thispagestyle{empty}

\begin{flushright}
ROM2F/2012/04 \\

\end{flushright}

\vspace{1.5cm}

\begin{center}

{\LARGE {\bf 
Correlation Functions of 
Conserved \\  Currents in  Four Dimensional  \\
Conformal Field Theory \rule{0pt}{25pt} }} \\
\vspace{1cm} \ { Yassen~S.~Stanev 
} \\
\vspace{0.6cm}  {{\it I.N.F.N.\ -\ Sezione
di Roma \ ``Tor Vergata''}} \\ {{\it Via della Ricerca  Scientifica, 1}}
\\ {{\it 00133 \ Roma, \ ITALY}}
\end{center}

\vspace{1cm}

\begin{abstract}

We derive a generating function for all the 3-point functions of higher spin conserved currents in four dimensional conformal field theory. The resulting expressions have a rather surprising factorized form which suggest that they can all be realized by currents built from free massless fields of arbitrary  (half-)integer spin $s$. This property is however not necessarily true also for the higher-point functions. As an illustration we analyze the general 4-point function of conserved abelian $U(1)$ currents of scale dimension equal to three and find that apart from the two free field realizations there is a unique possible function which may correspond to an interacting theory. Although this function passes several non-trivial consistency tests, it remains an open challenging problem whether it can be  actually realized in an interacting CFT. 

\end{abstract}

\vspace{2cm}
\noindent
\rule{6.5cm}{0.4pt}
\begin{flushleft}
$^{\dag}$ Yassen.Stanev@roma2.infn.it
\end{flushleft}

\end{titlepage}

\vfill
\newpage

\section{Introduction. Summary of the results}
\label{sec:INTRO}

In the last decade most of the research in Conformal Field Theory (CFT) in $D=4$ dimensional space time was performed in the framework of supersymmetric gauge theories, like  ${\cal N}=4$  SYM to mention just one prominent example.
This was paralleled by a continuous improvement in the understanding of the properties of the conformal partial wave expansions, initially only for  external scalar fields \cite{Partialwaves}, but recently also some progress 
for the general case was made \cite{newPartialsym,newPartial}.
A quite different approach, based on the notion of Global Conformal Invariance 
(GCI) \cite{NT}, and  bilocal conformal fields was followed in \cite{Rational1,Rational2,BNRT}.
In particular, in this setup it was proven that 
a GCI theory always contains an infinite number of 
conserved symmetric traceless tensors. Moreover, if the algebra of observables is generated by  a hermitian scalar field $\Phi_2$ of scale dimension two, then all the functions can be realized in terms of massless scalar fields.

Recently, in \cite{MZ}, in $D=3$ dimensions a stronger result has been proven, 
namely that all the correlation functions of the observables  in any CFT  
with higher spin symmetry (\ie   with an infinite number of 
conserved symmetric traceless tensors) are equal to the ones in a free field theory. This can be viewed as an extension of the famous 
Coleman-Mandula theorem \cite{Colem-M} to the case of CFT, where there is no  $S-$matrix.
It is natural to ask whether one can establish a similar result also in $D=4$ dimensional CFT.

In this paper, 
as a first step in this direction we study the 3-point functions of conserved symmetric tensors in $D=4$ dimensions. 
The main results of our analysis can be summarized as follows: 
\begin{itemize}
\item We derive a very compact generating function for all the 3-point functions of conserved symmetric tensors, eqs.(\ref{generatingFeven},\ref{generatingFodd}). 
\item 
The set of 3-point functions  can be organized in families, which exhibit a rather surprising factorization property, eqs.(\ref{Gfacteven1}, \ref{Gfacteven2}), namely  
 the functions in each family have a common factor $G(r,r,r)$, which is again a conserved 3-point function.
\end{itemize}
The factorization property leads us to conjecture that the functions in each family can be realized by currents built from free massless fields of arbitrary  half-integer spin $s=r/2$.

Even if all the 3-point functions admit a free field construction, 
higher-point functions are not necessarily free. As an illustration we analyze the general rational 4-point function of conserved abelian $U(1)$ currents of scale dimension equal to three and find that apart from the two free field realizations
(massless complex scalar and massless Weyl fermion)  there is a unique possible function which may correspond to an interacting theory, eq.(\ref{simple}). 
We study in details the properties of this solution and show that it satisfies 
several non-trivial conditions. The proof of the positivity of this function remains an open challenging problem.

The paper is organized as follows, in Section~2 we review some general properties of the higher spin conserved currents and their correlation functions
in $D=4$ dimensional CFT.
In Section~3 we derive a generating function for all the conformal invariant 3-point functions of the higher spin conserved currents.  Section~4 is devoted to the analysis of the 4-point conformal invariant correlation functions of four abelian $U(1)$ currents. Finally, in Section~5 we give our conclusions.

\section{Higher spin conserved currents in CFT}
\label{sec:HSCC}

In this section we shall study  the 3-point correlation functions of conserved currents. Before proceeding let us briefly recall some relevant properties of conserved currents in $D=4$ dimensional CFT and introduce some useful notation. 
Consider a symmetric traceless tensor $J_{(\mu_1 \dots \mu_r)}(x)$ of rank $r$, which belongs to the representation $(r/2,r/2)$ of the Lorentz group and has scale dimension $\Delta_r$. 
The conservation condition 
$\partial^{\mu_1} J_{(\mu_1 \dots \mu_r)}(x) = 0 $ is conformal invariant only if the scale dimension  $\Delta_r$ of the current is related to its rank $r$ by $\Delta_r=r+2$. The quantity $\Delta_r-r$ is called the twist of the field, so all the conserved currents have twist two. The $r=0$ case of twist two operator is special, it corresponds to a scalar field of scale dimension equal to two which does not obey a conservation condition, but it shares many properties with the family of conserved currents. The $r=1$ field is the usual
dimension three current $J_{\mu}(x)$, while the $r=2$ field is the dimension  four symmetric traceless stress-energy tensor $\Theta_{\mu \nu}(x)$, which generates conformal transformations. With a slight abuse of notation,  
we shall call a conserved higher spin current any (twist two) symmetric traceless tensor $J_{(\mu_1 \dots \mu_r)}(x)$ of rank $r \geq 0$, which belongs to the representation $(r/2,r/2)$ of the Lorentz group and has scale dimension $\Delta_r=r+2$. 
One may contract the indices of $J_{(\mu_1 \dots \mu_r)}(x)$ with an auxiliary light-like vector $a$,  defining 
\be
J_r(a,x) \ = \ a^{\mu_1} \dots a^{\mu_r} J_{\mu_1 \dots \mu_r}(x) \ , \quad  {\rm where} \ \  a^2=0 \ .
\label{Ja}
\ee
Note that with this notation both the symmetry and the tracelessness are automatic. 
The transformation law for $J_r(a,x)$ under local conformal transformations is
\be 
\left[ C_{\alpha} , J_r(a,x) \right] = 
(2 x_\alpha (x.\partial_x + \Delta_r) -x^2 \partial_{x^\alpha} +
2 \, a_\alpha \, x.\partial_a -2 \, a.x \, \partial_{a^\alpha})  J_r(a,x) , 
\label{conformalJ}
\ee 
and the conservation condition can be written as \cite{BargTod}
\be
\left((a.\partial_a +1) \ \partial_a.\partial_x -\frac{1}{2}~ a.\partial_x  \ \Box_a \right) J_r(a,x) \ = \ 0 .
\label{consa}
\ee
The conformal invariant 2-point function of two currents $J_r$ is given by 
\be
\langle 0 | \  J_{r_1}(a,x_1) \ J_{r_2}(b,x_2) \ | 0 \rangle \ = \ \delta_{r_1 r_2}  \, c(r_1) \, \frac{(R_{ab})^{r_1}}{x_{12}^4} \ , 
\label{2point}
\ee
where $c(r)$ is a constant, $x_{ij}=x_i-x_j$, and $R_{ab}$ is the 2-point conformal covariant of weights $(1,1)$ in $x_1$ and $x_2$ 
\be 
R_{ab} = R_{ab}(x_{12}) \ = \ \frac{1}{x_{12}^2}\left( {a.b - 2 \ \frac{a.x_{12} \ b.x_{12}}{x_{12}^2}}\right) \ .
\label{defR}
\ee  
Let us stress that the compact form of the 2-point function of  eq.~(\ref{2point}) is due to the light-like conditions on the auxiliary vectors $a^2=b^2=0$. If these are not taken into account, for $r_i \geq 2$  in the right hand side of eq.~(\ref{2point}) 
there are various trace subtraction terms proportional to $a^2b^2$. We shall comment more on this after presenting the general form of the 3-point function  
\be
G(r_1,r_2,r_3) \ = \ \langle 0 | \  J_{r_1}(a,x_1) \ J_{r_2}(b,x_2) \  J_{r_3}(c,x_3) \ | 0 \rangle \ ,
\label{3point}
\ee
where without loss of generality we can choose $r_1 \leq r_2 \leq r_3$.

There are exactly seven independent primitive conformal covariants which can be made out of 3-points in $D=4$ dimensions. Any other covariant can be written as a linear combination of products of the primitive ones.
One involves the completely antisymmetric pseudotensor  $\epsilon_{\mu \nu \rho \tau }$ and is odd under parity. It is given by a rather complicated expression  (see eq.~(\ref{oddinv}) below). The remaining six ones \cite{Me88} are even under parity and we shall list them together with their conformal weights in $(x_1,x_2,x_3)$. Three are essentially 2-point ones
obtained from (\ref{defR}) by permutations
\ba 
 R_{ab} &=&  R_{ab}(x_{12}) \ \  {\rm of \ weights} \ (1,1,0) \ , \nn \\
 R_{ac} &=&  R_{ac}(x_{13}) \ \ {\rm of \ weights} \ (1,0,1) \ , \nn \\
 R_{bc} &=&  R_{bc}(x_{23}) \ \ {\rm of \ weights} \ (0,1,1) \ . 
\label{R3pt}
\ea
The other three are genuine 3-point covariants 
\ba 
L_{a} &=& L_{a}(123) =  \frac{a.x_{12}}{x_{12}^2} - \frac{a.x_{13}}{x_{13}^2} \ \ {\rm of \ weights} \ (1,0,0) \ , \nn \\
L_{b} &=& L_{b}(231) =  \frac{b.x_{23}}{x_{23}^2} - \frac{b.x_{21}}{x_{12}^2} \ \ {\rm of \ weights} \ (0,1,0) \ , \nn \\
L_{c} &=& L_{c}(312) =  \frac{c.x_{31}}{x_{13}^2} - \frac{c.x_{32}}{x_{23}^2} \ \ {\rm of \ weights} \ (0,0,1) \ . 
\label{L3pt}
\ea
It is easy to show that the above six covariants form a complete set of even covariants. Indeed one can make use of conformal invariance and send one of the points $x_i$ in eq.~(\ref{3point}) to infinity. The  prescription for doing it is 
\be 
J_{(\mu_1 \dots \mu_r)}(\infty) \ = \ \lim_{x \rightarrow \infty} x^{2 \Delta_r}
\prod_{i=1}^r \left(\delta^{\nu_i}_{\mu_i}-2 \, \frac{x_{\mu_i} x^{\nu_i}}{x^2} \right)  J_{(\nu_1 \dots \nu_r)}(x) \ . 
\label{pointtoinf}
\ee 
If we send $x_1$ to infinity, we can still use translation invariance to send $x_3$ to zero, then the 3-point function eq.~(\ref{3point}), will depend only on $a,b,c$ and $z=x_{23}$, which gives precisely six possible structures, namely $a.b$, $a.c$, $b.c$, $a.z$, $b.z$ and $c.z$ (since $a^2=b^2=c^2=0$, and the powers of $z^2$ are determined by the scaling properties).  
The same argument proves also that the odd covariant is unique, since out of four vectors one can construct only one odd structure, namely $a^{\mu} b^{\nu} c^{\rho} z^{\tau} \epsilon_{\mu \nu \rho \tau }$. We shall choose the odd covariant to be of weights $(1,1,1)$ in $(x_1,x_2,x_3)$. This convention follows the one for the even covariants above and allows a uniform treatment of all the three point functions, since the weights are always equal to the power of the relative auxiliary light-like vectors, hence the total weights of a product of covariants 
will be equal to the ranks of the respective currents. 
The odd covariant can be written in several different ways,   
we shall use the following form, which is manifestly invariant under simultaneous permutations of $a$ and $b$ and $x_1$ and $x_2$ etc.
\ba 
&& O_{abc} = \frac {1}{x_{12}^2 x_{13}^2 x_{23}^2} \left[ 2\,(a.x_{12}+ a.x_{13}) \, \epsilon^{b,c,x_{12},x_{13}} + 2 \, (b.x_{12} - b.x_{23}) \,  \epsilon^{a,c,x_{12},x_{23}}  \right.  \nn \\
&&  \left.  - 2 (c.x_{13}+ c.x_{23}) \, \epsilon^{a,b,x_{13},x_{23}} 
 + x_{12}^2  \epsilon^{a,b,c,x_{12}} - x_{13}^2  \epsilon^{a,b,c,x_{13}} 
+ x_{23}^2  \epsilon^{a,b,c,x_{23}}
\right] ,
\label{oddinv}
\ea  
where  $ \epsilon^{a,b,c,z} = a^{\mu} b^{\nu} c^{\rho} z^{\tau} \epsilon_{\mu \nu \rho \tau }$. 

Note that the trick of sending a point to infinity can be used also the other way around. In other words, given the scale dimensions of the three currents, any 
Lorentz and scale covariant function of $z=x_{23}$ can be uniquely reconstructed 
to a conformal 3-point function. 

In $D=4$ space-time dimensions the covariants $R$ and $L$ eqs.(\ref{R3pt},\ref{L3pt}) are independent, any even power of the odd one $O_{abc}$ eq.(\ref{oddinv}) can be expressed in terms of the  even ones, while (since the odd one is unique) any  odd power 
of $O_{abc}$ can be written as  a product of $O_{abc}$ with an even covariant\footnote{Let us note that in $D=3$ dimensions the situation is completely different, since  there is a non-linear relation between the even covariants \cite{Osborn1}, and there are three distinct odd covariants 
\cite{Giombi}. Note also that in $D \geq 5$ dimensions, there are no odd covariants, simply because there are no enough independent vectors to contract the indices of the antisymmetric tensor $\epsilon$.}.
We can summarize these considerations as follows: 

The general form of the parity even conformal invariant 3-point function of higher spin currents eq.(\ref{3point}) is a linear combination of monomials of the form 
\be   
G(r_1,r_2,r_3)\mid_{\rm \, even} \,  = 
  \frac{1}{x_{12}^2 x_{13}^2 x_{23}^2} \sum C_{k_1,k_2,k_3,n_1,n_2,n_3} \, R_{ab}^{k_1} R_{ac}^{k_2} R_{bc}^{k_3} L_a^{n_1} L_b^{n_2} L_c^{n_3}  \ , 
\label{geneven}
\ee    
with $k_1+k_2+n_1=r_1$, $k_1+k_3+n_2=r_2$, $k_2+k_3+n_3=r_3$. 

The general form of the parity odd conformal invariant 3-point function of higher spin currents eq.(\ref{3point}) is a linear combination of monomials of the form 
\be   
G(r_1,r_2,r_3)\mid_{\rm \, odd} \, = 
  \frac{O_{abc}}{x_{12}^2 x_{13}^2 x_{23}^2} \sum C_{k_1,k_2,k_3,n_1,n_2,n_3} \, R_{ab}^{k_1} R_{ac}^{k_2} R_{bc}^{k_3} L_a^{n_1} L_b^{n_2} L_c^{n_3}  \ , 
\label{genodd}
\ee    
with $k_1+k_2+n_1=r_1-1$, $k_1+k_3+n_2=r_2-1$, $k_2+k_3+n_3=r_3-1$.

The constants $C_{k_1,k_2,k_3,n_1,n_2,n_3}$ can be determined by imposing the conservation conditions eq.(\ref{consa}). The universal form of the prefactor is due to our conventions for the weights of the primitive covariants, and is determined by the twist(=2) of the currents. 

Before proceeding, let  us briefly comment on some extensions and limitations of our formalism. Although we have written eqs.(\ref{geneven},\ref{genodd}) for the case of symmetric tensors of twist two, their generalization to the case of three tensors of well defined parity, but with different symmetry and scale dimensions is straightforward. The covariants $R,L$ and $O$ still form a basis for these 3-point functions and the only necessary modifications are in the powers of the prefactor, and possibly in the use of a more sophisticated set of auxiliary vectors to reflect the symmetry of the tensors. If however one wants to describe spinors, spin-tensors, but also more complicated integer spin representations of the Lorentz group which are not eigenstates of parity (\eg (2,0)), then our list of covariants is clearly not sufficient. An important progress in deriving the necessary building blocks in the general case has been made in \cite{TodLambda},
where a systematic use  is made of two component auxiliary spinors (instead of vectors) to contract the Lorentz indices of the fields. In particular, this allows to introduce new primitive 2-point invariants $P_{ij}$
appearing in the functions involving fermions, as well as some new 3-point invariants appearing in functions of selfdual antisymmetric tensors. 
In this setup, the 2-point covariant $R_{ab}$ is not primitive anymore, rather it is bilinear in $P$,  $R_{ab} = 2 P_{12} P_{21}$.   
Another important difference between the notation used in \cite{TodLambda} and our conventions, which is manifest also for the conserved higher spin currents, is that the auxiliary spinors transform non-trivially
under scale and conformal transformations. This remains true also for the auxiliary vectors (called $\zeta$), which are bilinear in the spinors, while our auxiliary vectors $a,b,c$ are inert under scale and special conformal transformations.
Thus although the 2- and 3-point expressions in eqs.(\ref{R3pt},\ref{L3pt}) formally are the same, in the notation of \cite{TodLambda} they are called invariants, while in our terminology they are covariants of certain weights
in $(x_1,x_2,x_3)$.

\section{Generating function for the 3-point functions}
\label{sec:Generating}

In this section we shall derive a generating function for all the 3-point functions $G(r_1,r_2,r_3)$, eq.(\ref{3point}), of the higher spin conserved currents $J_{r}(a,x)$ (\ref{Ja}). The first question to answer is: How many 
independent 3-point functions are there for given fixed $r_1, r_2, r_3$ ?
Apart from the correlation functions present in the three free massless field theories
(scalar, Weyl fermion and Maxwell field), which give a lower bound  for the number of 3-point functions of conserved currents, 
the answer in $D=4$ dimensions was explicitly known only in a limited number of cases for 
 functions of small rank ($r_1,r_2 \leq 2$) currents. In particular there are
two even and one odd 3-point functions for
three currents ($r_1=r_2=r_3=1$) \cite{Schreier}. There is a unique
odd 3-point function of an axial current ($r_1=1$) and two stress tensors ($r_2=r_3=2$) \cite{Erdmenger}.
There are exactly two even 3-point functions involving two currents ($r_1=r_2=1$) and a stress tensor ($r_3=2$), while  there are exactly three (even) 3-point functions involving three stress tensors ($r_1=r_2=r_3=2$)  \cite{Me88}. The latter result has been derived also in \cite{Osborn1}, for arbitrary space time dimension $D \geq 4$,
and generalized for arbitrary $r_3$ in \cite{counting}, where also a conjecture 
for the total number of invariants was made.
Let us note that for $D=3$, there are generically only two even 3-point 
functions \cite{Osborn1,Giombi} for any rank of the tensors. 

In order to determine the number of 3-point functions we imposed the conservation conditions eq.(\ref{consa}) on the general even/odd conformal  invariant 3-point functions of eq.(\ref{geneven}) / eq.(\ref{genodd}) for fixed 
$r_i$ and solved
the systems of equations for the constants $C_{k_1,k_2,k_3,n_1,n_2,n_3}$.
Sending one point to infinity with the help of eq.(\ref{pointtoinf}) drastically 
simplifies the computations. The result of this (exhaustive for $r_i \leq 50 $) study can be summarized as follows

\begin{itemize}
\item The number of independent even conserved 3-point functions is 
\be 
{\rm Number}( \, G(r_1,r_2,r_3)\mid_{\rm \, even}) \ = \  {\rm min}(r_1,r_2,r_3)+1 \ ;
\label{numbereven}
\ee 
\item The number of independent odd conserved 3-point functions is
\be 
{\rm Number}( \, G(r_1,r_2,r_3)\mid_{\rm \, odd})  \ = \ {\rm min}(r_1,r_2,r_3) \ ;
\label{numberodd}
\ee 
\end{itemize}
and is in perfect agreement with the numbers  conjectured in \cite{counting}.
Although our rather experimental approach cannot substitute a rigorous proof, we 
are convinced that the above relations are correct beyond any reasonable doubt, and shall assume them in the rest of the paper. 
These numbers can be understood as follows\footnote{I'm very grateful to Ivan Todorov for suggesting this interpretation.}. Apart from the three Lagrangian free field theories mentioned above in $D=4$ dimensions one can consider also free massless fields transforming under the representation $(s,0)$, $s$ (half-)integer $>1$ (and its complex conjugate $(0,s)$) of the Lorentz group. Out of these fields one can construct only conserved tensors of rank $r \geq 2 s$. Taking into account 
all these 3-point functions, one gets an agreement with the numbers in eqs.(\ref{numbereven},\ref{numberodd}).  

Before proceeding let us note that counting the 3-point functions for generic space time dimension $D \geq 5$ gives the same number of  even 3-point functions 
eq.(\ref{numbereven}) (we recall that there are no odd 3-point 
functions in  $D \geq 5$).

In order to analyze the constraints of conformal symmetry on the 4- and higher  point functions (for example by exploiting like in \cite{MZ} the action of the conserved charges arising from the higher spin conserved currents) one needs an explicit construction of all the different 3-point functions. This task is considerably more difficult in $D=4$ than for $D=3$, since an exhaustive case by case analysis of the free field theories is not possible. Instead we shall derive a generating function ${\cal G}$ for all the 3-point functions, with the following property  
\begin{itemize}
\item ${\cal G}$ generates the correct number of linear independent, conformal invariant, conserved in all three arguments, 3-point functions.
\end{itemize}

Since the derivation is rather long, we shall first present the final expressions.
For the even 3-point functions 
\be   
{\cal G}\mid_{\rm \, even} \ = \ {1 \over x_{12}^2 x_{13}^2 x_{23}^2} \,
 {w \over v} \, 
  {w-{1 \over 2} \, \zeta \, X_{abc} 
 \over w^2-\zeta \, X_{abc} \, w  -{1 \over 2} \, \zeta^2 \, R_{ab} R_{ac} R_{bc} }  \ , 
\label{generatingFeven}
\ee
while for the odd 3-point functions
\be   
{\cal G}\mid_{\rm \, odd} \ = \ {1 \over x_{12}^2 x_{13}^2 x_{23}^2} \, 
 {w \over v} \,
  {\zeta \, O_{abc} 
 \over w^2-\zeta \, X_{abc} \, w  -{1 \over 2} \, \zeta^2 \, R_{ab} R_{ac} R_{bc} }  \ , 
\label{generatingFodd}
\ee
where 
\ba 
X_{abc} &=& \frac{}{} 2 \, L_a L_b L_c + R_{ab} L_c + R_{ac} L_b + R_{bc} L_a \ , \nn \\
u &=& \frac{}{} 2 \, (1-L_a) (1-L_b) (1-L_c) +  \nn \\ 
&&    \frac{}{} + R_{ab} (1-L_c)+R_{ac} (1-L_b)+R_{bc} (1-L_a) \ , \nn \\
v &=& \frac{}{} \sqrt{ u^2 +2 \, R_{ab} R_{ac} R_{bc}} \ , \nn \\
w &=& \frac{}{} u+v \ , 
\label{uvw}
\ea  
and $R$, $L$, and $O$ are the primitive covariants defined in eqs.(\ref{R3pt},\ref{L3pt},\ref{oddinv}). 

The above expressions generate the 3-point functions of the conserved higher spin currents in the following way. 
Let us rescale 
\ba 
&& L_a \longrightarrow  \rho_1 L_a \, , \qquad 
 L_b \longrightarrow  \rho_2 L_b \, , \qquad 
L_c \longrightarrow  \rho_3 L_c , \nn \\
&& R_{ab} \rightarrow \rho_1 \rho_2 R_{ab} , \quad  
 R_{ac} \rightarrow \rho_1 \rho_3 R_{ac} ,  \quad  
R_{bc} \rightarrow \rho_2 \rho_3 R_{bc} , \nn \\
&&  O_{abc} \rightarrow \rho_1 \rho_2 \rho_3 O_{abc} ,
\label{rescaleLR}
\ea 
which implies $X_{abc} \rightarrow \rho_1 \rho_2 \rho_3 X_{abc}$,
and expand eq.(\ref{generatingFeven}) / (\ref{generatingFodd}) in power series in $\rho_i$ and $\zeta$.  
Then the coefficients of $\rho_1^{r_1} \rho_2^{r_2} \rho_3^{r_3} \zeta^n$
are even / odd 3-point functions for three higher spin conserved currents of ranks 
$r_1, r_2, r_3$, for any value of $n$ from 0 / 1 up to min($r_1,r_2,r_3$). 
The upper limit, which is in accord with eqs.(\ref{numbereven},\ref{numberodd}),  is due to the fact that after the rescaling (\ref{rescaleLR}), 
 in eqs.(\ref{generatingFeven},\ref{generatingFodd}) $\zeta$ is always multiplied by  $\rho_1 \rho_2 \rho_3$.

The proof of the conservation is easier working with the compact expressions for the generating functions. Applying 
the differential operator (\ref{consa}) on the functions (\ref{generatingFeven},\ref{generatingFodd}), after very long but straightforward algebra one obtains vanishing result.

The striking similarity of the generating functions for the even and the odd 
3-point functions suggests that the separation in even/odd parts is somewhat  artificial.
One can consider instead a pair of related by parity generating functions 
\be 
{\cal G}^{\pm} \ = \ {\cal G}\mid_{\rm \, even} \pm \, \lambda \ {\cal G}\mid_{\rm \, odd} \ .
\label{complexG}
\ee   
However, we prefer to work with the even and the odd functions because they have well defined permutation symmetry. Under the simultaneous exchange of $(a, x_1)$ with $(b, x_2)$, the primitive covariants $L$, $R$ and $O$  change as follows
\ba 
&& L_a \rightarrow - L_b , \quad 
 L_b \rightarrow - L_a , \quad 
L_c \rightarrow - L_c , \nn \\
&& R_{ab} \rightarrow  R_{ab} , \quad  
 R_{ac} \rightarrow  R_{bc} ,  \quad  
R_{bc} \rightarrow  R_{ac} , \nn \\
&&  O_{abc} \rightarrow  O_{abc} ,
\label{permute12}
\ea 
and similarly under the other permutations ($L$ change sign, while $R$ and $O$ do not). Then it follows by inspection of eqs.(\ref{geneven}, \ref{genodd}) 
and comparing the parity of $n_1+n_2+n_3$  to the parity of $r_1+r_2+r_3$, that
whenever two of the $r_i$ are equal the even functions are symmetric/ antisymmetric under their permutation  if $r_1+r_2+r_3$ is even/ odd respectively, while the odd functions are symmetric/ antisymmetric under permutation if $r_1+r_2+r_3$ is odd/ even respectively.
In particular this means that the 3-point functions of three  equal currents of rank $r$ is necessarily of the even/odd type for $r$ even/odd. 

Another remarkable property (which is somewhat obscure in eqs.(\ref{generatingFeven},\ref{generatingFodd})) is that the 3-point functions can be written in a factorized form. To be explicit we shall spell out this for the case of the even 3-point functions, but the same considerations are valid also for the odd ones. Let us introduce the auxiliary conformal covariants $I_r$ of weights $(r,r,r)$ in $(x_1,x_2,x_3)$ defined by the expansion 
\be   
  {1-{1 \over 2} \, z \, X_{abc} 
 \over 1-z \, X_{abc}   -{1 \over 2} \, z^2 \, R_{ab} R_{ac} R_{bc} }  \ = \ \sum_r   \, z^r \, I_r \ , 
\label{Specialeven}
\ee
where $X_{abc}$ is defined in (\ref{uvw}).
The first terms in the expansion are $I_0=1$,  
\ba 
I_1 &=& \frac{1}{2} \,  X_{abc}     \ , \nn \\
I_2 &=&   \frac{1}{2} \, ( R_{ab} R_{ac} R_{bc} +  X_{abc}^2)  \ , \nn \\ 
I_3 &=&   \frac{1}{4} \,  X_{abc} \, ( 3  R_{ab} R_{ac} R_{bc} + 2 X_{abc}^2)  \  , \nn \\
I_4 &=&  \frac{1}{4} \, ( ( R_{ab} R_{ac} R_{bc})^2 + 4  R_{ab} R_{ac} R_{bc} X_{abc}^2 + 2 X_{abc}^4)   \ ,  \\ 
I_5 &=&  \frac{1}{8} \,  X_{abc} \, ( 5( R_{ab} R_{ac} R_{bc})^2 + 10  R_{ab} R_{ac} R_{bc} X_{abc}^2 + 4 X_{abc}^4)    \ , \nn \\ 
I_6 &=&  \frac{1}{8} \, ( R_{ab} R_{ac} R_{bc} +  X_{abc}^2) \, ( ( R_{ab} R_{ac} R_{bc})^2 + 8  R_{ab} R_{ac} R_{bc} X_{abc}^2 + 4 X_{abc}^4)   \ . \nn
\label{exampleIr}
\ea 
It is straightforward to verify that for any $r$ 
\be 
G(r,r,r) \ = \ \frac{I_r}{x_{12}^2 x_{13}^2 x_{23}^2}
\label{3prrr}
\ee
is an even  conserved conformal invariant function for three currents of equal rank $r$.   
One can rewrite eq.(\ref{generatingFeven})  as 
\be   
{\cal G}\mid_{\rm \, even} \ = \ \sum_r \, \zeta^r \, {\cal G}_r \ ,  
\label{Gfacteven1}
\ee 
where 
\be   
{\cal G}_r \ = \ \frac{G(r,r,r)}{ v \, w^r} \ ,  
\label{Gfacteven2}
\ee 
and $v$, $w$ are defined in eqs.(\ref{uvw}). Note that for any fixed $r$ the expansion of the denominator of eq.(\ref{Gfacteven2}) in power series in $L$ and $R$, gives rise to an infinite number of conserved 3-point functions with $r_i \geq r$. Hence we can organize the set of 3-point functions in families, 
such that the functions in each family have a common factor $G(r,r,r)$, which is again a conserved 3-point function.

We conjecture that this factorization property is due to the free field construction in terms of $(s,0)$ and $(0,s)$ fields. 
Indeed in each such model, the lowest rank conserved current corresponds to (an appropriate contraction of) the product of two free fields  and has rank $r=2s$, while all the other ones contain also derivatives and have higher rank, so the general picture is in agreement with the structure of the family ${\cal G}_{2s}$
in eq.(\ref{Gfacteven2}). We checked  that ${\cal G}_{0}$, ${\cal G}_{1}$ and  ${\cal G}_{2}$ generate the 3-point functions of conserved currents in the theories of scalar, fermion and Maxwell fields respectively, but we were not able to prove the conjecture in general.

Let us stress that both the factorization property and the (relative) simplicity of the generating functions  crucially depend on our choice to work with the conserved currents $J_r(a,x)$ of eq.(\ref{Ja})  with Lorentz indices contracted by light-like vectors $a$, such that $a^2=0$, since the trace  subtraction terms  do not respect the factorization.     

We shall give a brief description of the procedure we followed to derive the generating functions eqs.(\ref{generatingFeven}, \ref{generatingFodd}). The reason for this is two-fold. On one hand, 
since most of the 3-point functions of the higher spin conserved currents have no intrinsic normalizations (a notable exception are the functions of the stress-energy tensor which generates conformal transformations), there is a huge freedom in the explicit form of the generating functions,    
so we will motivate and illustrate some the choices we made to fix this freedom. On the other hand we believe that some of the intermediate formulae are of interest on their own. We shall concentrate on the even generating function, the procedure for the odd one is essentially the same.

 From the general form of eq.(\ref{geneven}), it follows that any conformal 3-point function (for fixed $r_i$) can be naturally  obtained by acting on $L_a^{r_1} L_b^{r_2} L_c^{r_3}$, with a suitable polynomial of the differential operators $D_{ab} = R_{ab} \partial_{L_a} \partial_{L_b}$, and similar for $D_{ac}$ and $D_{bc}$
\be   
G(r_1,r_2,r_3)\mid_{\rm \, even}   \ = \ 
  \frac{1}{x_{12}^2 x_{13}^2 x_{23}^2} \,  P(D_{ab}, D_{ac}, D_{bc}) 
 \, L_a^{r_1} L_b^{r_2} L_c^{r_3}  \ . 
\label{Polyd}
\ee  
The conservation conditions eq.(\ref{consa}) impose constraints on the coefficients of the polynomial.
With a bit of luck, a lot of patience and a fast computer, we found that 
for any $(r_1,r_2,r_3)$ and for any integer $0 \leq r \leq $ min$(r_1,r_2,r_3)$ the conformal invariant function 
\be 
G(r_1,r_2,r_3 ; r) \ = \ G(r,r,r) \, 
f_{ab}^{(r)} f_{ac}^{(r)} f_{bc}^{(r)}  \, (L_a^{r_1-r} L_b^{r_2-r} L_c^{r_3-r}) \  
\label{Gdiff}
\ee
is conserved. 
Here $G(r,r,r)$ is the conserved 3-point function defined in eq.(\ref{3prrr}), while $f_{ab}^{(r)}$ is the differential operator
\be 
f_{ab}^{(r)} \ = \ \sum_{k=0}^{\infty}{1 \over k! (k+r)!}\left(-{ R_{ab} \over 2}\right)^k
 {\partial^k \over \partial L_a^k} \, {\partial^k \over \partial L_b^k} \ ,
\label{Bessel}
\ee 
(which can be expressed also as a Bessel function), 
and similar for $f_{ac}^{(r)}$ and $f_{bc}^{(r)}$. Comparison with 
eq.(\ref{numbereven}) shows that the functions (\ref{Gdiff}) exhaust all the 3-point functions of conserved higher spin currents.
Apart from the already discussed factorization of $G(r,r,r)$, eq.(\ref{Gdiff})
has another surprising feature, namely the factorization of the three differential operators. We were not able to find a simple explanation of why this happens.

A generating function for $G(r_1,r_2,r_3; r)$ (for fixed value of $r$) can be obtained by summing them up with arbitrary non-vanishing coefficients $N_{(r_1,r_2,r_3)}^{(r)}$
\ba  
{\cal G}_{r} &=&  \sum_{r_1,r_2,r_3=r}^{\infty} N_{(r_1,r_2,r_3)}^{(r)} \, 
G(r_1,r_2,r_3; r) \nn \\
&=&  G(r,r,r) \,   f_{ab}^{(r)} f_{ac}^{(r)} f_{bc}^{(r)} \left( \sum_{r_1,r_2,r_3=r}^{\infty} N_{(r_1,r_2,r_3)}^{(r)} \,
 L_a^{r_1-r} L_b^{r_2-r} L_c^{r_3-r} \right) \, .
\label{genarb}
\ea
One may use this freedom, to get more compact expressions. We find 
it convenient to choose $N_{(r_1,r_2,r_3)}^{(r)}$ of the factorized form 
\be 
N_{(r_1,r_2,r_3)}^{(r)} = \frac{r_1!}{(r_1-r)! \, r!} \, 
 \frac{r_2!}{(r_2-r)! \, r!} \,  \frac{r_3!}{(r_3-r)! \, r!}  \, n(r) \ , 
\ee  
corresponding to  
\be 
\sum_{r_1,r_2,r_3=r}^{\infty} N_{(r_1,r_2,r_3)}^{(r)} \, L_a^{r_1-r} L_b^{r_2-r} L_c^{r_3-r}
 =  {n(r) \over [(1-L_a) (1-L_b) (1-L_c)]^{r+1}} \ .
\label{}
\ee 
This choice is motivated by the simplicity of the final formulae and is the result of a large number of attempts to simplify the action of the differential operators, guided by the $r=0$ case (corresponding to all $N_{(r_1,r_2,r_3)}^{(0)}=1$). After quite a lot of work, the 
evaluation of the differential operators and the resummation gives (up to an 
irrelevant numerical factor which can be compensated by appropriately choosing $n(r)$)
\be 
 f_{ab}^{(r)}f_{ac}^{(r)} f_{bc}^{(r)} 
\left( {1 \over [(1-L_a) (1-L_b) (1-L_c)]^{r+1}} \right) \ \propto \
 { 1 \over v \, w^r } \ , 
\ee
with $v$ and $w$ defined in eq.(\ref{uvw}). Putting everything together, 
we recover eq.(\ref{Gfacteven2}), and (with the help of eqs.(\ref{Specialeven},\ref{3prrr}) also the generating function for all the even 3-point functions  eq.(\ref{generatingFeven}).  Repeating exactly the same steps one derives also 
the generating function for all the odd 3-point functions  eq.(\ref{generatingFodd}).

\section{Four point function of conserved $U(1)$ currents}
\label{sec:4-point}

In this section we shall study the conformal invariant 4-point function of an abelian $U(1)$ current in $D=4$  dimensional CFT.  We shall restrict our attention to the very special class 
of rational functions, which arise naturally in the framework of GCI CFT \cite{Rational1}, and in theories with an infinite number 
of conserved higher spin currents. Recently it has been shown that such theories are necessarily free in three dimensions \cite{MZ}. 
The situation in four dimensions is less clear. It has been shown that the theories involving a scalar of dimension $\Delta=2$ are free \cite{Rational2,BNRT}, but in general not much is known. The reason for this is two-fold. On one hand, as we have seen the structure of the conformal 3-point 
functions, and hence also the Operator Product Expansion (OPE) structure, is more complicated in $D=4$ (compared to $D=3$). On the other hand, in order
to impose the positivity constraints on the 4-point functions (see below) 
one needs the conformal partial wave expansion, which, although some recent 
progress \cite{newPartialsym,newPartial},  so far have been worked out in all details only for scalar fields \cite{Partialwaves}. 
 
We start by  recalling some of the implications of conformal invariance on 4-point functions. Any even  conformal invariant 4-point function of twist two currents can be schematically written in the form 
\be
\frac{1} {x_{13}^4 x_{24}^4} \, \sum_i P_i(\{L,R\}) \, f_i(s,t) \ ,
\label{general4p}
\ee 
where $P_i(\{L,R\})$ are appropriate polynomials of the primitive covariants
$L$ and $R$, 
and $f_i(s,t)$ are functions of the two conformal invariant cross-ratios
\be 
s \ = \ \frac{x_{12}^2 x_{34}^2} {x_{13}^2 x_{24}^2} \ , \qquad 
t \ = \ \frac{x_{14}^2 x_{23}^2} {x_{13}^2 x_{24}^2} \ .
\label{st}
\ee 
As anticipated in (\ref{general4p}), all the even primitive covariants are effectively 2- and 3-point ones (see eqs.(\ref{R3pt},\ref{L3pt})), thus one has six different $R$s, and 
twelve $L$s (out of which only eight are independent, since 
$L_a(123)+L_a(134)=L_a(124)$ etc.). 
In particular, 
we will be interested in  the conformal 4-point functions of four $U(1)$   currents $J^{\alpha_i}(x_i)$, of scale dimension $\Delta_J=3$, satisfying the conservation condition
\be 
\partial_{\alpha} J^{\alpha}(x) \  = \ 0 \ . 
\label{J1cons} 
\ee
Note that since the current has rank one, there is no need to introduce the auxiliary light-like vectors, so one can use this simple conservation law, 
instead of (\ref{consa}). 

We shall first list all our assumptions and will briefly comment on their implications
\begin{itemize}

\item Even, conformal invariant function. Hence in  (\ref{general4p}) 
the polynomials $P_i$ are of  the form $LLLL$,  $LLR$ or $RR$. 
Note that the 4-point function of four equal fields is always even. 

\item Full permutation symmetry. This relates different terms in (\ref{general4p}) in  a rather nontrivial way, since the $L$s are not independent.

\item Conservation. This leads to conditions both on $P_i$ and on $f_i$. 

\item Weak positivity. Only unitary representations of the conformal group appear in the  OPE 
expansion of two currents. We recall that unitarity gives lower bounds on the 
scale dimensions of the fields. In particular, if a field belongs to
the irreducible representation $(j_1, j_2)$ of the Lorentz group, its scale dimension $\Delta$  has to satisfy \cite{Mack}
\ba 
\Delta &\geq& j_1+j_2+1 \quad {\rm if} \quad j_1 j_2=0 \ , \nn \\
\Delta &\geq& j_1+j_2+2 \quad {\rm if} \quad j_1 j_2 \neq 0 \ .
\label{unitary}
\ea  
This condition constrains the short-distance singularities of the function for small $x_{ij}$. 
Weak positivity is a necessary, but by far not sufficient condition to ensure
the complete positivity condition, which requires also that all the coefficients in the conformal partial wave expansion of the function are positive. We shall comment more on the positivity condition after presenting the solutions we found.

\item Rationality of the functions $f_i$. Combined with weak positivity, this means that all the $f_i$ in eq.(\ref{general4p}) are polynomials in $s$, $1/s$, $t$ and $1/t$ (of degree not higher than two for the function of four currents,
since the truncated part has to be strictly less singular than the 2-point functions). 
This is a very strong requirement indeed. In particular, it excludes from our analysis all the theories like ${\cal N}=4$ SYM, where the fields have non-integer anomalous dimensions. However, as we already mentioned, it is natural in the theories with higher spin symmetries. Let us note also that even a negative result, \ie that there are only the free field solutions to our assumptions, 
is of interest in this case. 
\end{itemize}

The main result of our analysis can be stated in the following way. There are only three different solutions of all these conditions. Two of them are realized
in free field theories, namely the free complex scalar field and the free massless Weyl fermion (we shall refer to them for brevity as the scalar and the fermion case in the rest of the paper). The third function, which we shall call the new one, does not correspond to a free field realization and can be written in the following compact way 
\ba 
\langle J^{\alpha_1}(x_1) J^{\alpha_2}(x_2) 
J^{\alpha_3}(x_3) J^{\alpha_4}(x_4)\rangle   = \
\epsilon^{\alpha_1\, \mu_1\, \nu_1\, \rho_1} 
\epsilon^{\alpha_2\, \mu_2\, \nu_2\, \rho_2} 
\epsilon^{\alpha_3\, \mu_3\, \nu_3\, \rho_3} 
\epsilon^{\alpha_4\, \mu_4\, \nu_4\, \rho_4}  \frac{}{}  \nn \\  
\times  
R_{\mu_1 \mu_2} (x_{12})  R_{\nu_1 \nu_3} (x_{13})  R_{\rho_1 \rho_4} (x_{14}) 
R_{\rho_2 \rho_3} (x_{23})  R_{\nu_2 \nu_4} (x_{24})  R_{\mu_3 \mu_4} (x_{34})  \frac{}{} \, ,
\label{simple}
\ea 
where $\epsilon$ is the totally antisymmetric tensor and $R$ is the 2-point primitive covariant defined in eq.(\ref{defR}).

It should be noted, that eq.(\ref{simple}) is the result of a long simplification. Originally, the function was obtained starting from the ansatz in eq.(\ref{general4p}) and imposing all the conditions. The result was a horrible expression with 145 terms, which we transformed to the compact form (\ref{simple}) using twice the well known identity  $  {\rm det}( \delta) = \epsilon \times \epsilon$. In this form both the  permutation symmetry and the conservation of the function are manifest, if one uses that $R_{\mu \nu}(x) = 1/2 \, \partial_{\mu} \partial_{\nu} {\rm ln}(x^2)$. Let us stress that, since there are terms 
with denominators  like  $x_{12}^2 x_{13}^2 x_{14}^2$, this function cannot be realized by conserved currents which are bilinear in some fundamental fields (as  in all free field theories). It is easy to determine the leading short distance  singularity, say for $x_{12}$ small, of this 4-point function.  Indeed, since only $R(x_{12})$ is divergent in this limit, the function behaves not worse than $1/x_{12}^2$ or $x_{12}^\mu x_{12}^\nu / x_{12}^4$, the latter giving also the leading light-cone singularity $1/x_{12}^4$. With some more work one can obtain the leading terms in the current-current OPE  consistent with eq.(\ref{simple}) 
\ba   
z^4 : J^{\mu}(x_1) \, J^{\nu}(x_2) :  &=&   z_\rho \, z_\tau \, 
{\cal A}^{([\mu,\rho],[\nu,\tau])}(x_2)  \frac{}{} \nn \\
 &+&  \left( z^{\mu} z_\rho \, \delta^\nu_\tau 
+z^{\nu} z_\rho  \, \delta^\mu_\tau
- \eta^{\mu\nu} z_\rho z_\tau \right) \Theta^{\rho \tau}(x_2)  \frac{}{} \nn \\
 &+&  \left( z^{\mu} z^{\nu} 
+ \frac{\eta^{\mu\nu}}{2} \, z^2 \right) \Phi_4(x_2) \ + \ O(z^3)  \frac{}{} \, , 
\label{OPE}
\ea 
where $z=x_{12}$.
Here $\Theta^{\rho \tau}$ is the symmetric traceless stress-energy tensor, 
$\Phi_4$ is a scalar field of scale dimension equal to four, while 
${\cal A}^{([\mu,\rho],[\nu,\tau])}$ is a tensor field of scale dimension equal to four, belonging to the (reducible)  representation $(2,0)\oplus(0,2)$ of the Lorentz group.
It is 
antisymmetric in $\mu$, $\rho$ and $\nu$, $\tau$, symmetric under the exchange
of the two pairs, traceless in any pair of indices and satisfies  
$\epsilon_{\mu \nu \rho \tau}{\cal A}^{([\mu,\rho],[\nu,\tau])}=0$.

The stress-tensor $\Theta^{\rho \tau}$ is the lowest dimensional representative of the infinite family of twist two symmetric tensors in the OPE, while $\Phi_4$ 
is the lowest dimensional instance of the higher (\ie larger than two) twist operators. In much the same way ${\cal A}^{([\mu,\rho],[\nu,\tau])}$ is the first representative of an infinite family of tensors of scale dimension $4+n$ 
belonging to the representation $(2+n/2,n/2)\oplus(n/2,2+n/2)$ of the Lorentz group.
Comparison with the unitarity conditions eq.(\ref{unitary}) shows that the operators in this family (for $n \neq 0$), as well as all the twist two operators, saturate  exactly the unitarity bound, while 
${\cal A}^{([\mu,\rho],[\nu,\tau])}$, $\Phi_4$ and all the higher twist operators have scale dimensions above the unitarity bound. It follows that the whole function (\ref{simple}) satisfies the weak positivity condition.

A natural question at this point is: What about complete positivity? In fact one can prove that the contributions of all the twist two even rank symmetric tensors to the function (\ref{simple}) respect positivity.
We recall that there are exactly two different 3-point functions (cf. eq.(\ref{numbereven}) for $r_1=r_2=1$)  of two $U(1)$ conserved currents
and a higher spin conserved current of rank $r$.
Since the leading light-cone singularity in the function (\ref{simple}) is $1/x_{12}^4$, it follows that all these contributions are proportional to the ones in the fermion case (i.e. to the ones in the theory of   free fermions, obtained from ${\cal G}_1$ in eq.(\ref{Gfacteven2})). 
To make this explicit, let us expand  ${\cal G}_1$ in power series and keep only the terms of order $r$ in $c$, linear in $a$ and $b$. 
They correspond to the  3-point functions of two currents with a rank $r$ conserved tensor and because of the factorization property have the simple form
\be 
G(1,1,r)_{\psi} \  \propto \  \frac{X_{abc} \, L_c^{r-1}}{x_{12}^2 x_{13}^2 x_{23}^2} \ , 
\label{Jrpsi}
\ee
where $X_{abc}$ is defined in eq.(\ref{uvw}). It is straightforward to check that the leading light-cone singularity of all these expressions is equal to
$1/x_{12}^4$. 
On the other hand, in the scalar case all such functions (obtained from ${\cal G}_0$ in eq.(\ref{Gfacteven2})) have a different leading light-cone singularity. Indeed, the linear in $a$ and $b$ part of ${\cal G}_0$
has the form 
\ba  
{\cal G}_0 |_{r1=r2=1} &\propto& \frac{1}{{x_{12}^2 x_{13}^2 x_{23}^2} \ (1-L_c)^3} \, 
\left(  2 \, L_a L_b - R_{ab} \frac{}{}  \right. \nn \\
&-&  4 \, L_a L_b L_c + 2 \, R_{ab} L_c - R_{bc} L_a -R_{ac} L_b \frac{}{}  \nn \\
 &+& \left.  2 \, L_a L_b L_c^2 - R_{ab} L_c^2 + L_a L_c R_{bc} + L_b L_c R_{ac} + R_{ac} R_{bc}
 \frac{}{}\right) .
\label{Jrphi}
\ea 
It follows that in this case all the functions $G(1,1,r)_{\varphi}$ have a stronger leading light-cone singularity, namely $1/x_{12}^6$ (in fact this is true separately for the contributions in the three lines of eq.~(\ref{Jrphi})),
hence these expressions cannot appear in the expansion of the function (\ref{simple}). 

  As we will now demonstrate, the contributions of the even rank conserved tensors to the function (\ref{simple})  
are not just proportional (which leaves open the question of the normalization constants, which could potentially violate positivity), rather they are all exactly equal to the ones in the fermion case, hence satisfy the positivity condition.  
To this end let us note that contracting the OPE of two currents appropriately
\be 
 V(x_1,x_2) \ = \ \left[ \, (x_{12})_\mu (x_{12})_\nu  : J^\mu(x_1) \, J^\nu(x_2) : - 
\, \frac{x_{12}^2}{2} : J^\mu(x_1) \, J_\mu(x_2) : \, \right] \ ,
\label{bilocal}
\ee 
we can project on the contributions of only the even rank symmetric tensors\footnote{Since only the even rank tensors contribute to the symmetric part of the OPE (see the discussion after eq.(\ref{permute12})).}.  
In fact this is true separately for both terms in the right hand side of 
eq.(\ref{bilocal}), but only the particular combination $V(x_1,x_2)$ has also nice 
conformal properties, namely it is a scalar conformal bi-field of weights $(2,2)$ in $x_1$ and $ x_2$ respectively. Note that although $V(x_1,x_2)$ in general receives contributions also from the higher twist operators, the scalar field $\Phi_4$ in the right hand side of eq.(\ref{OPE}) is projected out. Thus the only operator of scale dimension four in $V(x_1,x_2)$ is the stress-energy tensor $\Theta^{\rho \tau}$.   
Given the 4-point function of four currents one can compute the 4-point function  $\langle V(x_1,x_2) \, J^\rho(x_3) \, J^\tau(x_4)\rangle$.
Let us denote by $f^{\rho \tau}(x_1,x_2,x_3,x_4)$ the coefficient to $1/x_{12}^2$ in this function
\be 
f^{\rho \tau}(x_1,x_2,x_3,x_4) \ = \ \langle V(x_1,x_2) \, J^\rho(x_3) \,  J^\tau(x_4)\rangle \vert_\frac{1}{x_{12}^2}  \ . 
\label{twist2}
\ee 
It follows that $f$ contains the contributions of only the twist two  
even rank symmetric tensors, i.e. the even rank conserved higher spin currents in the OPE of two currents.
A rather long, but straightforward calculation shows that 
\be 
f^{\rho \tau}(x_1,x_2,x_3,x_4)_{\rm new} \ = \ c \, f^{\rho \tau}(x_1,x_2,x_3,x_4)_{\psi} \ , 
\label{new_psi}
\ee 
where the subscript "new" denotes $f$ computed from the 
function (\ref{simple}), the subscript $\psi$ denotes $f$ computed from 
the current 4-point function for a free fermion and $c$ is a positive constant. Since the right hand side is computed in a unitary free field model and respects positivity, it follows that 
the contributions of all the even rank conserved tensors in (\ref{simple})
also satisfy positivity.
Let us stress that eq.(\ref{new_psi}) does not mean that the contributions of all conserved tensors in the two functions coincide.
Indeed, in the free fermion case there is also the family of odd rank 
conserved tensors, which contributes to the antisymmetric under the exchange of $\mu$ and $\nu$ part of the OPE, hence is not captured by $V$ (and $f$). Rather, eq.(\ref{new_psi}) states that the contributions of the operators which appear in the expansion of  both functions are the same. This has a very important consequence, namely any linear combination
of the function (\ref{simple}) and the free fermion current 4-point function will admit an OPE involving only one conserved tensor of each rank. 
In particular there will be only one operator with the quantum numbers of the stress-energy tensor. This suggests that eq.(\ref{simple}) is better suited as an interaction contribution for the fermion case. 

Let us illustrate this with an example. 
Assume that one starts with a 4-point function 
which is a linear combination of the function in the free complex scalar case  and (\ref{simple}) (or the free fermion function).
As already explained, the light cone singularities of the terms coming from the stress-energy tensor in these two functions are different, namely $1/x_{12}^6$ and $1/x_{12}^4$ respectively. Thus, performing the double expansion in $x_{12}$ and $x_{34}$, and keeping track only of the leading light-cone singularities of the different terms, one finds for the contribution of the  stress-energy tensor $\Theta^{\rho \tau}$ 
\be  
 \langle JJJJ \rangle_\varphi + \langle JJJJ \rangle_{\rm new}   
 \vert_{\Theta} \ = \ 
\left( \frac{c_\varphi^2}{x_{12}^6 x_{34}^6}
+ \frac{c_n^2}{x_{12}^4 x_{34}^4} \right) \langle \Theta \Theta \rangle \ . 
\label{scsp}
\ee 
But this expression cannot be reproduced by any OPE involving only one tensor. 
One needs a current-current OPE of the form
\be  
 J(x_1) \, J(x_2) \ = \ \left(\frac{c_\varphi}{x_{12}^6}+\frac{c_n}{x_{12}^4}\right) T_1
+ \left(\frac{c_\varphi}{x_{12}^6}-\frac{c_n}{x_{12}^4}\right) T_2 + \dots ,  
\label{T1T2}
\ee 
where $T_1$ and $T_2$ are two distinct (orthogonal) symmetric traceless tensors with 2-point functions   
\be
 \langle T_1 T_2 \rangle = 0 , \quad \langle T_1 T_1 \rangle  = \langle T_2 T_2 \rangle = \frac{1}{2} \, \langle \Theta \Theta \rangle \ .
\ee 
Hence, a function like (\ref{scsp}) contradicts the uniqueness of the operator with the quantum numbers of the stress-energy tensor. Let us note that this is a rather strong requirement. Already in the case of a free complex scalar field $\varphi$ there are three such operators, the stress-energy tensor built from $\varphi$ and $\varphi^{*}$ and two more made from two $\varphi$ or two $\varphi^{*}$. However, this proliferation of the conserved currents, specially when they appear simultaneously in the same OPE, is usually a signal that the
theory is a direct sum of two or more simpler theories. 

In contrast, starting with a linear combination of the function in the  free fermion case and of (\ref{simple}), due to eq.(\ref{new_psi}), one finds 
\be  
 \langle JJJJ \rangle_\psi + \langle JJJJ \rangle_{\rm new} 
\vert_{\Theta} \ = \ 
\frac{c_\psi^2+c_n^2}{x_{12}^4 x_{34}^4} \  \langle \Theta \Theta \rangle \ , 
\label{sp_new}
\ee  
and similar for all the other even rank conserved tensors. Clearly, such a function is reproduced by   
an OPE involving only one conserved tensor of each rank. 

Note also that in the OPE of two stress-energy tensors in the theory of free
Maxwell field $F^{\mu \nu}$ naturally appears the operator\footnote{The last 
(proportional to $\epsilon$) 
term, being totally antisymmetric, gives a vanishing contribution to the OPE of two stress-energy tensors, but ensures that the operator in eq.(\ref{FF}) belongs to the representation $(2,0)\oplus(0,2)$ of the Lorentz group.}
\ba  
F^{\mu \nu} F^{\alpha \beta} 
&+& \frac{1}{2} \left( 
\eta^{\mu \alpha} {F^{\nu}}_\rho F^{\rho \beta}
-\eta^{\nu \alpha} {F^{\mu}}_\rho F^{\rho \beta}
-\eta^{\mu \beta} {F^{\nu}}_\rho F^{\rho \alpha}
+\eta^{\nu \beta} {F^{\mu}}_\rho F^{\rho \alpha}
\right) \nn \\
&-& \frac{1}{6} \left( 
\eta^{\mu \alpha} \eta^{\nu \beta}-\eta^{\mu \beta}\eta^{\nu \alpha} \right) F_{\tau \rho} F^{\rho \tau} 
+ \frac{1}{24} \,
\epsilon^{\mu \nu \alpha \beta} \epsilon_{\kappa \lambda \rho \tau } \, 
F^{\kappa \lambda} F^{\rho \tau}  \ , 
\label{FF}
\ea 
which has all the properties of  ${\cal A}^{([\mu, \nu],[\alpha, \beta])}$ in eq.(\ref{OPE}).
We believe that this is not accidental and may help to solve the first of the two remaining unsolved problems, namely to check  
positivity for the family starting with ${\cal A}$, 
as well as for the higher twist contributions in the function (\ref{simple}).

\section{Conclusions}
\label{sec:CONO}

In this paper we 
derived a generating function for all the 3-point functions of conserved higher 
spin currents in $D=4$ dimensional CFT. The particular factorized form of the resulting expressions suggests that they can all be realized in terms of free fields. In order to test whether this is the case also for higher point functions, we studied the rational conformal invariant 4-point functions of four abelian $U(1)$ currents and found only one function which cannot occur in a free field theory, eq.(\ref{simple}). Although we were not able to prove the positivity of this function, there are some indications that (if consistent) it may correspond to the interaction of massless fermions with (possibly non-abelian) gauge fields.
On one hand, any linear combination of the function  eq.(\ref{simple}) and the 4-point function of four $U(1)$ currents in the theory of free massless fermions is consistent with the assumption of uniqueness of the  stress-energy tensor, see eq.(\ref{new_psi}). On the other hand the  operator ${\cal A}^{([\mu,\rho],[\nu,\tau])}$, appearing in the OPE of two currents eq.(\ref{OPE}), has the same quantum numbers as the operator in eq.(\ref{FF}),
which naturally arises in the gauge theory. A preliminary analysis of the 
rational 4-point functions of the stress-tensor in $D=4$ CFT,
gives some additional support to this conjecture.

Another interesting open problem is to construct higher point functions of the 
currents consistent with (\ref{simple}). Let us recall, that the $(n+2)$-point functions can be related to the $n$-point ones by a two step procedure. Indeed, performing the OPE, one first derives the $(n+1)$-point function of the stress-energy tensor and $n$ currents, which in turn is related to the $n$-point function of the currents by the Ward identities. Both steps lead to  
non-trivial constraints, so already the existence of a 6-point function with the desired properties is not granted. For this analysis a construction of the current $J^{\alpha}(x)$ as a composite in terms of elementary fields could be very helpful. This goes beyond the scope of this paper,
let us only note that the structure of eq.(\ref{simple}) suggests a cubic ansatz.

\section*{Acknowledgements}

\noindent
It is a pleasure to thank Ivan Todorov for
numerous illuminating discussions and for showing me his manuscript \cite{TodLambda} prior to publication. 
This work was supported in
part by MIUR-PRIN contract 2009-KHZKRX-005.

\end{document}